\pdfoutput=1
\documentclass[preprint,aps,showpacs,showkeys,nofootinbib]{revtex4}
\usepackage{amsmath}
\usepackage{graphicx}
\usepackage{booktabs}
\usepackage{epstopdf}
\usepackage{slashed}
\usepackage{mathrsfs}
\usepackage{diagbox}
\usepackage{array}
\usepackage{color}
\begin{document}
\title{S, T, U Parameters in The B-LSSM}
\author{Sheng-Kai Cui$^{a}$\footnote{email:cuisk261@163.com},Ke-Sheng Sun$^{b}$\footnote{email:sunkesheng@126.com},Yu-Li Yan$^{c}$\footnote{email:yychanghe@sina.com},Jin-Lei Yang$^{c,d,e}$\footnote{email:jlyang@hbu.edu.cn},Tai-Fu Feng$^{a,c,d,e,f}$\footnote{email:fengtf@hbu.edu.cn}}
\affiliation{$^a$Department of Physics, Guangxi University, Nanning, 530004, China}
\affiliation{$^b$Department of Physics, Baoding University, Baoding 071000, China}
\affiliation{$^c$Department of Physics, Hebei University, Baoding 071002, China}
\affiliation{$^d$Hebei Key Laboratory of High-precision Computation and Application of Quantum Field Theory, Baoding, 071002, China}
\affiliation{$^e$Hebei Research Center of the Basic Discipline for Computational Physics, Baoding, 071002, China}
\affiliation{$^f$ Department of Physics, Chongqing University, Chongqing, 401331, China}
\begin{abstract}
Using the pinch technique, we compute the one-loop vertices of weak interactions in the B-LSSM and incorporate their pinch contributions into the gauge boson self-energies. Compared to the definitions of the $S$, $T$, and $U$ parameters in the Standard Model based on the $SU(2)_L \otimes U(1)_Y$ group, the corresponding parameters in the local B-L gauge symmetry (B-LSSM) are modified. We provide these redefined $S$, $T$, and $U$ parameters and demonstrate the convergence of the results. In the framework of the low-energy effective Lagrangian for weak interactions, the $S$, $T$, and $U$ parameters can be expressed as functions of certain parameters in the B-LSSM. The updated experimental and fitting results constrain the parameter space of the B-LSSM strongly.
\end{abstract}


\maketitle

\section{Introduction\label{INT}}
\indent\indent
The measurement results for the mass of the $W$ boson, as reported by the ATLAS experimental group in 2023 \cite{ATLAS:2023fsi}, are in close agreement with the theoretical predictions of the Standard Model (SM). This high-precision agreement provides a stringent benchmark for probing theories beyond the SM. In particular, models with extended gauge sectors, such as the B-LSSM, must accommodate this result, which can lead to significant constraints on their parameter space. In comparison to the SM, the inclusion of the additional $U(1)_{B-L}$ gauge group in the B-LSSM leads to the emergence of a more substantial $Z'$ boson, which in turn modifies the original definition of the Weinberg angle. The electroweak radiative corrections of observable quantities can be obtained through the oblique parameter method, and the superfields in the B-LSSM exert a notable influence on these oblique parameters. In this paper, we derive the gauge-invariant gauge boson self-energies in the B-LSSM. Based on experimental results, we impose constraints on the parameters of the model.

The B-LSSM \cite{Ambroso:2010pe,FileviezPerez:2010ek,Barger:2008wn,FileviezPerez:2008sx,Yang:2020bmh,Yang:2021duj,Yang:2018fvw,Yang:2018guw,Yang:2018utw,Zhang:2021nzv,Yang:2019aao,Yang:2020ebs,Dong:2021cxn,Dong:2020ioc,Dong:2024lvs,Yang:2023krd,Cui:2020nju} is based on the gauge symmetry group $SU(3)_C \otimes SU(2)_L \otimes U(1)_Y \otimes U(1)_{B-L}$, where B represents the baryon number and L represents the lepton number. The B-LSSM not only provides an explanation for the existence of a small mass for left-handed neutrinos but also offers a solution to the little hierarchy problem \cite{Abdallah:2016vcn} in the Minimal Supersymmetric Standard Model (MSSM). The gauge invariance of $U(1)_{B-L}$ allows for the conservation of R-parity, which is typically assumed in the MSSM to prevent proton decay. When this symmetry is spontaneously broken, R-parity conservation can still be maintained \cite{Aulakh:1999cd}. This model helps to understand the origin of R-parity and the potential ways it could be broken in supersymmetric models. Additionally, compared to the MSSM, the B-LSSM provides more candidates for dark matter \cite{Khalil:2008ps,DelleRose:2017ukx}.

The $S$, $T$, and $U$ parameters \cite{Grimus:2008nb,Maksymyk:1993zm,Cacciapaglia:2004rb,Lavoura:1992np,Haywood:1999qg,Cacciapaglia:2004jz,Burgess:1993mg,Asadi:2022xiy,Long:1999bny,Pich:2013fea,Rehman:2025urc} present an extension of the method developed by Peskin \cite{Peskin:1991sw}, building on the work of Kennedy and Lynn \cite{Kennedy:1988td}, to handle radiative corrections in electroweak interaction processes. Although the B-LSSM introduces an additional $U(1)_{B-L}$ gauge group compared to the SM, the oblique parameter framework remains a powerful tool at the electroweak scale. The new gauge sector modifies the observables directly via tree-level gauge kinetic mixing and mass matrix deformations, which can be elegantly encoded into the effective $S$, $T$, and $U$ parameters.

 When calculating physical observables or considering physical processes, their gauge invariance is typically taken into account. It becomes critical when extracting physical information from S-matrix elements, as gauge invariance cannot always be guaranteed. The oblique parameters discussed in this paper must account for gauge invariance. To address this issue, the pinch technique (PT), a well-established method based on the background field method (BFM), is employed. This technique has found widespread adoption and application. For more details, see references \cite{Binosi:2009qm,Hashimoto:1994ct,Denner:1994xt,Denner:1994nn,Denner:1996wn,Denner:1995jd,Papavassiliou:1993ex,Degrassi:1992ue,Binosi:2002ft,Binosi:2003rr,Papavassiliou:1995gs,Binosi:2002ez,Papavassiliou:1996zn,Papavassiliou:1994pr}.

While many studies on oblique parameters rely heavily on Standard Model Effective Field Theory (SMEFT) global fits, it is important to recognize the conceptual distinction between these effective coefficients and the full theoretical calculations. To provide a more fundamental theoretical foundation, this work aims to avoid "over-fitting" the full theory directly to these effective parameters. Instead, we start from the formal theory of the B-LSSM to rigorously derive the gauge-invariant analytic expressions at the one-loop level, focusing on the physical trends and sensitivities revealed by the theoretical calculation itself.

Based on this robust framework, our numerical analysis strategy focuses on characterizing the dominant new physics sources. First, we systematically analyze how the mass mixing in the key supersymmetric sectors (namely the squark and neutralino-chargino sectors) influences the electroweak precision observables at the loop level. Second, for the additional $U(1)_{B-L}$ gauge parameters, we employ a complementary approach, utilizing the effective Lagrangian to impose direct constraints at the tree level. This dual approach lays the necessary methodological foundation and serves as a physical template for understanding how different symmetry-breaking sectors collectively impact low-energy precision measurements.

The structure of this paper is outlined as follows: In Section \ref{BL}, we provide a brief introduction to the B-LSSM and the construction of the $Z^{'}$ boson. In Section \ref{PT}, we present the gauge-invariant self-energies in the B-LSSM, derived using the pinch technique. In Section \ref{FOC}, we introduce the gauge-invariant $S$, $T$, and $U$ parameters in the B-LSSM, and based on these results, we further constrain the gauge mixing angle $s'$ by incorporating effective Lagrangian techniques.In Section \ref{dcns}, we proved the divergence cancellation of two sectors. In Section \ref{NA}, we analyze the experimental data's constraints on the parameter space. Finally, in Section \ref{CON}, we present our conclusions.

\section{The B-L SSM\label{BL}}
\indent\indent
The superpotential of the B-LSSM is given by
\begin{eqnarray}
&&W=Y_u^{ij}\hat{Q_i}\hat{H_2}\hat{U_j^c}+\mu \hat{H_1} \hat{H_2}-Y_d^{ij} \hat{Q_i} \hat{H_1} \hat{D_j^c}
-Y_e^{ij} \hat{L_i} \hat{H_1} \hat{E_j^c}+\nonumber\\
&&\;\;\;\;\;\;\;\;\;Y_{\nu, ij}\hat{L_i}\hat{H_2}\hat{\nu}^c_j-\mu' \hat{\eta}_1 \hat{\eta}_2
+Y_{x, ij} \hat{\nu}_i^c \hat{\eta}_1 \hat{\nu}_j^c,
\end{eqnarray}
where $i, j$ are generation indices, $\hat{\eta}_{1}\sim(1,1,0,-1)$, $\hat{\eta}_{2}\sim(1,1,0,1)$ are two chiral singlet superfields and $\hat{\nu}$ are three generations of right-handed neutrinos. The gauge group of $U(1)_{B-L}$ spontaneously broke without the simultaneous breaking of R-parity. The soft breaking terms of the B-LSSM are given by
\begin{eqnarray}
&&\mathcal{L}_{soft}=\Big[-\frac{1}{2}(M_1\tilde{\lambda}_{B} \tilde{\lambda}_{B}+M_2\tilde{\lambda}_{W} \tilde{\lambda}_{W}+M_3\tilde{\lambda}_{g} \tilde{\lambda}_{g}+2M_{BB'}\tilde{\lambda}_{B'} \tilde{\lambda}_{B}+M_{B'}\tilde{\lambda}_{B'} \tilde{\lambda}_{B'})-
\nonumber\\
&&\hspace{1.4cm}
B_\mu H_1H_2 -B_{\mu'}\tilde{\eta}_1 \tilde{\eta}_2 +T_{u,ij}\tilde{Q}_i\tilde{u}_j^cH_2+T_{d,ij}\tilde{Q}_i\tilde{d}_j^cH_1+
T_{e,ij}\tilde{L}_i\tilde{e}_j^cH_1+T_{\nu}^{ij} H_2 \tilde{\nu}_i^c \tilde{L}_j+\nonumber\\
&&\hspace{1.4cm}
T_x^{ij} \tilde{\eta}_1 \tilde{\nu}_i^c \tilde{\nu}_j^c+h.c.\Big]-m_{\tilde{\nu},ij}^2(\tilde{\nu}_i^c)^* \tilde{\nu}_j^c-
m_{\tilde{q},ij}^2\tilde{Q}_i^*\tilde{Q}_j-m_{\tilde{u},ij}^2(\tilde{u}_i^c)^*\tilde{u}_j^c-m_{\tilde{\eta}_1}^2 |\tilde{\eta}_1|^2-\nonumber\\
&&\hspace{1.4cm}
m_{\tilde{\eta}_2}^2 |\tilde{\eta}_2|^2-m_{\tilde{d},ij}^2(\tilde{d}_i^c)^*\tilde{d}_j^c-m_{\tilde{L},ij}^2\tilde{L}_i^*\tilde{L}_j-
m_{\tilde{e},ij}^2(\tilde{e}_i^c)^*\tilde{e}_j^c-m_{H_1}^2|H_1|^2-m_{H_2}^2|H_2|^2.
\end{eqnarray}
The terms with tilde denotes the supersymmetric partner of the corresponding chiral superfield. To obtain the masses of the physical neutral Higgs bosons, The Higgs fields are usually redefined as:
\begin{eqnarray}
&&H_1^1=\frac{1}{\sqrt2}(v_1+{\rm Re}H_1^1+i{\rm Im}H_1^1),
\qquad\; H_2^2=\frac{1}{\sqrt2}(v_2+{\rm Re}H_2^2+i{\rm Im}H_2^2),\nonumber\\
&&\tilde{\eta}_1=\frac{1}{\sqrt2}(u_1+{\rm Re}\tilde{\eta}_1+i{\rm Im}\tilde{\eta}_1),
\qquad\;\quad\;\tilde{\eta}_2=\frac{1}{\sqrt2}(u_2+i{\rm Re}\tilde{\eta}_2+i{\rm Im}\tilde{\eta}_2)\;.
\end{eqnarray}
In accordance with this definition, the majority of the symmetries pertaining to the $SU(2)_L\otimes U(1)_Y\otimes U(1)_{B-L}$ groups have been disrupted, with only the residual electromagnetic symmetry observed in the electromagnetic symmetry groups $U(1)_{B-L}$ remaining intact. The group $U(1)_{B-L}$ introduces new gauge boson $Z'$ and the corresponding gauge coupling constant $g_{_B}$. In addition, two Abelian groups give rise to a new effect absent in the MSSM or other SUSY models with just one Abelian gauge group: the gauge kinetic mixing. Here, we write the covariant derivative:
\begin{equation}
D_{\mu}=\partial_{\mu}-i
\begin{pmatrix}
    Y & Y_B
\end{pmatrix}
\begin{pmatrix}
    \widetilde{g} & \widetilde{g}_{YB} \\
    \widetilde{g}_{BY} & \widetilde{g}_B
\end{pmatrix}
\begin{pmatrix}
    \widetilde{A}_{\mu}^Y \\
    \widetilde{A}_{\mu}^B
\end{pmatrix},
\end{equation}
where $\widetilde{A}_{\mu}^Y$ and $\widetilde{A}_{\mu}^B$ denote the gauge fields associated with the two $U(1)$ gauge groups. $Y$ and $Y_B$ denote the hypercharge and $B-L$ charge respectively. It is easier to work with non-canonical covariant derivatives instead of off-diagonal field-strength tensors:
\begin{equation}
D_{\mu}=\partial_{\mu}-i
\begin{pmatrix}
    Y & Y_B
\end{pmatrix}
\begin{pmatrix}
    \widetilde{g} & \widetilde{g}_{YB} \\
    \widetilde{g}_{BY} & \widetilde{g}_B
\end{pmatrix}
R^T R
\begin{pmatrix}
    \widetilde{A}_{\mu}^Y \\
    \widetilde{A}_{\mu}^B
\end{pmatrix}.
\end{equation}
We insert the unitary matrix into the definition equation, which redefines both the coupling coefficient and gauge fields:
\begin{equation}
\begin{pmatrix}
    \widetilde{g} & \widetilde{g}_{YB} \\
    \widetilde{g}_{BY} & \widetilde{g}_B
\end{pmatrix} R^T =
\begin{pmatrix}
    g_1 & g_Y \\
    0 & g_B
\end{pmatrix}, \quad
R \begin{pmatrix}
    \widetilde{A}_{\mu}^Y \\
    \widetilde{A}_{\mu}^B
\end{pmatrix} =
\begin{pmatrix}
    A_{\mu}^Y \\
    A_{\mu}^B
\end{pmatrix}.
\end{equation}
Now we can give the mass matrix in the basis $(A^Y, B^3, A^B)$:
\begin{equation}
M^2_N = \frac{1}{8}
\begin{pmatrix}
    g_1^2 v^2 & g_1 g_2 v^2 & g_1 g_Y v^2 \\
    g_1 g_2 v^2 & g_2^2 v^2 & g_2 g_Y v^2 \\
    g_1 g_Y v^2 & g_2 g_Y v^2 & g_Y^2 v^2 + g_B^2 u^2
\end{pmatrix} \label{mn}.
\end{equation}
The mass matrix defined in Eq. (\ref{mn}) can be diagonalized by unitary matrix $R_N$:
\begin{equation}
R_N =
\begin{pmatrix}
    c_W & -s_W & 0 \\
    s_W & c_W & 0 \\
    0 & 0 & 1
\end{pmatrix}
\begin{pmatrix}
    1 & 0 & 0 \\
    0 & c' & -s' \\
    0 & s' & c'
\end{pmatrix} =
\begin{pmatrix}
    c_W & -s_W c' & s_W s' \\
    s_W & c_W c' & -c_W s' \\
    0 & s' & c'
\end{pmatrix},
\end{equation}
with
\begin{equation}
\begin{pmatrix}
    \gamma \\
    Z \\
    Z'
\end{pmatrix} = R_N^T
\begin{pmatrix}
    A^Y \\
    B^3 \\
    A^B
\end{pmatrix}.
\end{equation}
If we do some inverse operations on the mass matrix:
\begin{align}
&\frac{1}{4}
\begin{pmatrix}
    0 & 0 & 0 \\
    0 & G^2 v^2 & -G g_{YB} v^2 \\
    0 & -G g_{YB} v^2 & g_{YB}^2 v^2 + 4 g_B^2 u^2
\end{pmatrix} \nonumber \\
&=
\begin{pmatrix}
    1 & 0 & 0 \\
    0 & c' & -s' \\
    0 & s' & c'
\end{pmatrix}
\begin{pmatrix}
    0 & 0 & 0 \\
    0 & m_Z^2 & 0 \\
    0 & 0 & m_{Z'}^2
\end{pmatrix}
\begin{pmatrix}
    1 & 0 & 0 \\
    0 & c' & s' \\
    0 & -s' & c'
\end{pmatrix} \nonumber \\
&=
\begin{pmatrix}
    0 & 0 & 0 \\
    0 & {c'}^2 m_Z^2 + {s'}^2 m_{Z'}^2 & c' s' (m^2_Z - m^2_{Z'}) \\
    0 & c' s' (m^2_Z - m^2_{Z'}) & {s'}^2 m^2_Z + {c'}^2 m^2_{Z'}
\end{pmatrix},
\end{align}
where $G$ is electroweak coupling coefficient $G=\frac{e}{s_Wc_W}$, $v^2=v_1^2+v_2^2$ and $u^2=u_1^2+u_2^2$. Then, $g_B$ and $g_{YB}$ can be represented by some observable measurements:
\begin{align}
g_{YB}&=G\frac{c's'(m^2_Z-m^2_{Z'})}{{c'}^2m^2_Z+{s'}^2m^2_{Z'}},\nonumber\\
g_B&=\frac{G}{2}\frac{xm_Zm_{Z'}}{{c'}^2m^2_Z+{s'}^2m^2_{Z'}}.\label{gb}
\end{align}
Where $x=\frac{v}{u}$.

\section{Pinch Technique\label{PT}}

\indent\indent
The gauge-invariant self-energies of the $Z$ and $Z'$ bosons can be extracted from any neutral-current process. Following the approach of  Papavasiliou \cite{Papavassiliou:1994pr}, we consider the scattering process $e^+e^-\rightarrow \mu^+\mu^-$, The $S$-matrix element of this process should remain independent of the gauge-fixing parameters $\xi_i$ at any order of perturbative calculation. The propagators for the gauge bosons $\gamma$, $Z$ and $Z^{'}$ which include the gauge-fixing parameters, are given by:
\begin{eqnarray}
&&\Delta_{\mu\nu}{\gamma}(q)=\frac{-i}{q^2}[g_{\mu\nu}-(1-\xi_{\gamma})\frac{q_\mu q_\nu}{q^2}],\\
&&\Delta_{\mu\nu}Z(q)=\frac{-i}{q^2}[g_{\mu\nu}-(1-\xi_{Z})\frac{q_\mu q_\nu}{q^2-\xi_{Z}m^2_Z}],\\
&&\Delta_{\mu\nu}Z^{'}(q)=\frac{-i}{q^2}[g_{\mu\nu}-(1-\xi_{Z^{'}})\frac{q_\mu q_\nu}{q^2-\xi_{Z^{'}}m^2_{Z^{'}}}].
\label{GBP}
\end{eqnarray}
Traditionally, the $S$-matrix element $T$ can be decomposed into three parts based on the Mandelstam variables s, t and the external masses $m_e$, $m_{\mu}$. In terms of self-energy, vertex, and box contributions, $T_1$, $T_2$, and $T_3$ correspondingly contribute to the total expression:
\begin{eqnarray}
T(s,t,m_e,m_{\mu})=T_1(t)+T_2(t,m_e,m_{\mu})+T_3(s,t,m_e,m_{\mu})
\label{T}
\end{eqnarray}

\begin{figure}
\begin{minipage}[c]{1\textwidth}
\centering
\includegraphics[width=4.0in]{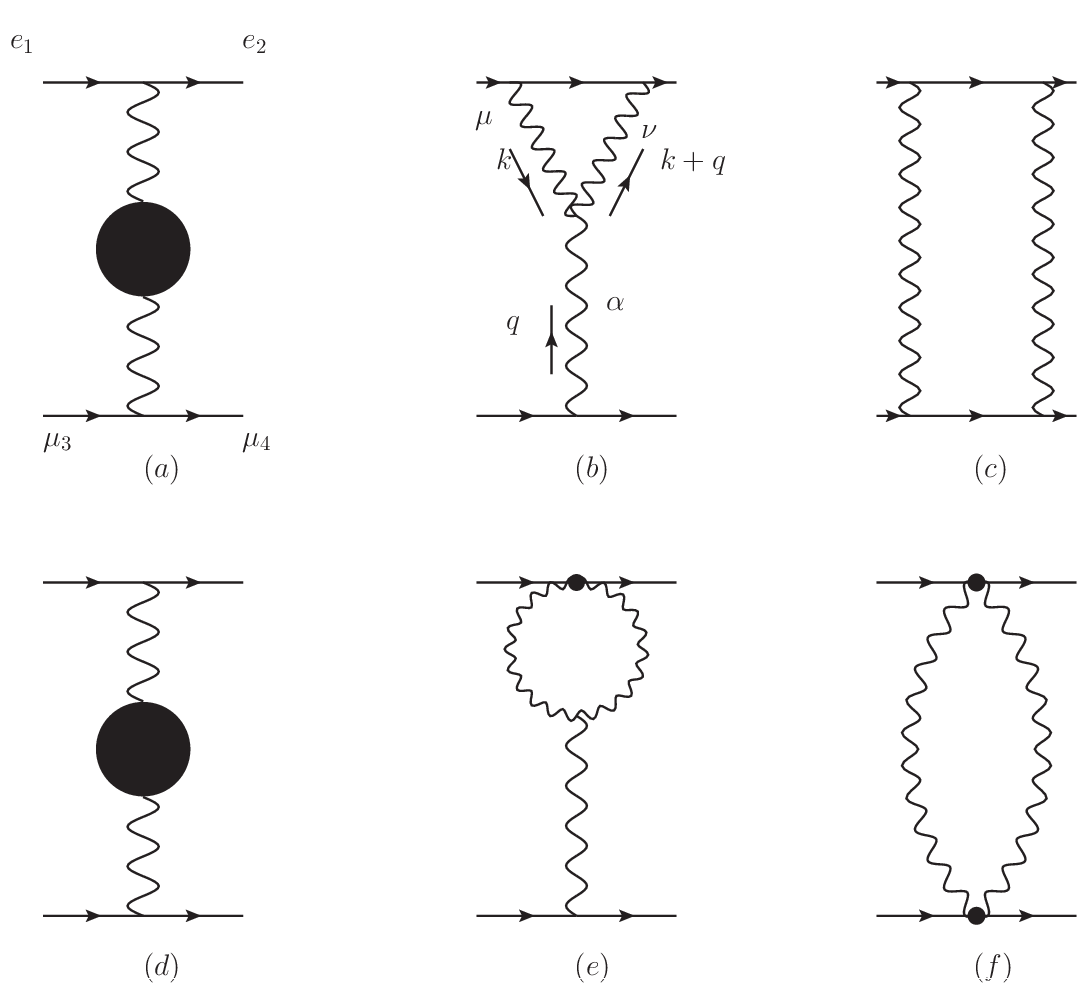}
\end{minipage}%
\caption[]{Diagram (a)-(c) show representative one-loop contributions to the S-matrix of the four-fermion process. Diagram (a) is the conventional gauge-dependent self-energy. The pinch parts extracted from diagrams (b) and (c)--shown separately in (e) and (f)--are systematically absorbed into (a), thereby converting it into the gauge-invariant effective self-energy displayed in (d).}
\label{psf}
\end{figure}

Although the total $T$ is gauge-independent, the individual components $T_1$ $T_2$ and $T_3$ are gauge dependent. By employing the pinch technique, we recast these into gauge-invariant quantities $\hat{T}_1$, $\hat{T}_2$, and $\hat{T}_3$:
\begin{eqnarray}
T(s,t,m_e,m_{\mu})=\hat{T}_1(t)+\hat{T}_2(t,m_e,m_{\mu})+\hat{T}_3(s,t,m_e,m_{\mu}).
\label{HAT}
\end{eqnarray}
From this, we derive the gauge-invariant self-energy $\hat{T}_1$.

The PT is a systematic approach. The form of the Ward identity changes depending on the specific physical process. For instance, when a gluon mediates strong interactions, the Ward identity takes the form:
\begin{align}
k^\mu\gamma_\mu\equiv\slashed{k}&=(\slashed{p}+\slashed{k}-m_i)-(\slashed{p}-m_i)\label{WIG}\\
&=S^{-1}_i(p+k)-S_i^{-1}(p)\nonumber.
\end{align}
For charged $W^{\pm}$ or neutral $Z$ bosons coupled to fermions, the Ward identity becomes:
\begin{align}
\label{WIW}k^\mu\gamma_\mu P_L\equiv\slashed{k}P_L&=(\slashed{p}+\slashed{k}-m_i)P_L-P_R(\slashed{p}-m_j)+m_iP_L-m_jP_R\\
&=S_i^{-1}(p+k)P_L-S_j^{-1}(p)P_R+m_iP_L-m_jP_R\nonumber.
\end{align}
Here, $P_{R,L}=\frac{1\pm\gamma_5}{2}$. The terms $S_i^{-1}(p+k)$ and $S_j^{-1}(p)$ vanish on shell, leaving behind $m_iP_L-m_jP_R$. When charged $W^{\pm}$ couples to fermions, the masses $m_i$ and $m_j$ differ. For the $Z$ boson coupling to a fermion-antifermion pair of equal mass, $m_i = m_j$.

Additionally, the vertex in Fig. \ref{psf}(b) is decomposed into two components:
\begin{eqnarray}
\Gamma_{\mu\nu\alpha}=\Gamma^{P\xi}_{\mu\nu\alpha}+\Gamma^\xi_{\mu\nu\alpha},
\label{VB}
\end{eqnarray}
where
\begin{eqnarray}
&&\Gamma^{P\xi}_{\mu\nu\alpha}=\frac{1}{\xi}[(q+k)_\nu g_{\mu\alpha}+k_\mu g_{\nu\alpha}] ,\\
&&\Gamma^{\xi}_{\mu\nu\alpha}=2q_\mu g_{\nu\alpha}-2q_\nu g_{\mu\alpha}-(2k+q)_\alpha g_{\mu\nu}+(1-\frac{1}{\xi})[k_\nu g_{\mu\alpha}+(k+q)_\mu g_{\nu\alpha}] .\nonumber
\label{VPF}
\end{eqnarray}
The term $\Gamma^{\xi}_{\mu\nu\alpha}$ satisfies the Ward identity
\begin{eqnarray}
q^\alpha\Gamma^{\xi}_{\mu\nu\alpha}=\Delta^{-1}_{\mu\nu}(k,\xi)-\Delta^{-1}_{\mu\nu}(k+q,\xi) .
\label{WIV}
\end{eqnarray}
Gauge boson propagators depend on the gauge-fixing parameters:
\begin{eqnarray}
\Delta^i_{\mu\nu}(q)=\mathscr{D}^i_{\mu\nu}(q)-\frac{q_\mu q_\nu}{M^2_i}D^i(q) ,
\label{FP}
\end{eqnarray}
where
\begin{eqnarray}
&&D^{i}(q,\xi)=\frac{i}{q^2-\xi_iM_i^2},\nonumber\\
&&\mathscr{D}^i_{\mu\nu}(q)=[g_{\mu\nu}-\frac{q_\mu q_\nu}{M^2_i}]\frac{-i}{q^2-M_i^2}.
\label{FP1}
\end{eqnarray}
$\mathscr{D}^i_{\mu\nu}(q)$ are the $W^{\pm}$,$Z$ and $Z^{'}$ propagators ($\xi\rightarrow\infty$). Selecting the Feynman gauge($\xi=1$) can greatly simplify calculations\cite{Papavassiliou:1994pr}.
Then, the pinch part of Fig .\ref{psf}(c) vanished.  Furthermore, when extracting the Pinch contribution of self energy, the following formula will be used:
\begin{eqnarray}
g^\alpha_\nu=&&\Delta^i_{\nu\mu}(q,\xi_i)\Delta^{-1}_i(q,\xi_i)^{\mu\alpha}\nonumber\\
=&&\Delta^i_{\nu\mu}(q,\xi_i)\mathscr{D}_i^{-1}(q)^{\mu\alpha}-iq_\nu q^\alpha D_i(q,\xi),\nonumber\\
q_\mu=&&-i\{q^2D_i(q,\xi_i)q_\mu+M_i^2q^\nu\Delta^i_{\nu\mu}(q,\xi_i)\},
\label{gq}
\end{eqnarray}
where
\begin{eqnarray}
\mathscr{D}_i^{-1}(q)^{\mu\alpha}=i\{g^{\mu\alpha}(q^2-M_i^2)-q^\mu q^\alpha\}.
\label{g1}
\end{eqnarray}
If $i=\gamma$,
\begin{align}
{\Delta^{\gamma}_{\mu\nu}}^{-1}(q)&=iq^2[g^{\mu\nu}+(1-\frac{1}{\xi_{\gamma}})\frac{q_{\mu}q_{\nu}}{q^2}].
\label{p}
\end{align}
In the B-LSSM, the pinch part of the self-energy of $W^{\pm}$ is similar to the result in the SM, here are the results:
 \begin{eqnarray}
\Pi^P_{WW}(q^2)=-4g_2^2(q^2-M_W^2)\{s_W^2I_{W\gamma}(q^2)+c_W^2({c'}^2I_{WZ}(q^2)+{s'}^2I_{WZ'}(q^2))\}.
\label{pw}
\end{eqnarray}
Where,
 \begin{eqnarray}
I_{ij}(q^2)=i\mu^{4-n}\int\frac{d^nk}{(2\pi)^n}\frac{1}{(k^2-m_i^2)[(k+q)^2-m^2_j]}.
\label{Ixx}
\end{eqnarray}
The calculation of guage-independent self-energies of $\gamma$,$Z$ and $Z^{'}$ needs to utilize the following allocation:
\begin{eqnarray}
\frac{g_2}{2}\gamma^\mu P_L=(-i)s_W(e\gamma e)^\mu-ic_Wc'(eZe)^\mu+ic_Ws'(eZ'e)^\mu,
\label{g3}
\end{eqnarray}
Where $(eZe)$ and $(eZ'e)$ denote the vertex of the corresponding particles. By combining Eq. (\ref{gq}) and Eq. (\ref{g3}), the following results can be obtained:
\begin{align}
&\Pi^P_{\gamma\gamma}(q^2)=-4g_2^2s_W^2q^2I_{WW}(q^2),\nonumber\\
&\Pi^P_{\gamma Z}(q^2)=-2g_2^2s_Wc_Wc^{'}(2q^2-M_Z^2)I_{WW}(q^2),\nonumber\\
&\Pi^P_{\gamma Z^{'}}(q^2)=-2g_2^2s_Wc_Ws^{'}(2q^2-M_{Z^{'}}^2)I_{WW}(q^2),\nonumber\\
&\Pi^P_{ZZ}(q^2)=-4g_2^2c^2_W{c^{'}}^2(q^2-M_Z^2)I_{WW}(q^2),\nonumber\\
&\Pi^P_{Z^{'}Z^{'}}(q^2)=-4g_2^2c^2_W{s^{'}}^2(q^2-M_{Z^{'}}^2)I_{WW}(q^2),\nonumber\\
&\Pi^P_{ZZ^{'}}(q^2)=2g_2^2c^2_Ws^{'}c^{'}(2q^2-M_Z^2-M_{Z^{'}}^2)I_{WW}(q^2).
\label{NCP}
\end{align}
By utilizing PT, we successfully derive gauge-invariant self-energies for the $W^{\pm}$, $Z$, $Z'$ and $\gamma$ bosons, which are essential for subsequent calculations.

\section{Formalism of Oblique Corrections\label{FOC}}
\indent\indent
\subsection{Formalism of S T U paramaters in the B-LSSM\label{formalism}}
The implementation of renormalization schemes that utilize the number of observables equivalent to the number of free parameters may prove challenging in practice. This is due to the necessity of solving a more extensive set of equations than those encountered in the SM, which may result in the generation of highly intricate and unwieldy analytical formulae \cite{Chankowski:2006jk}.

The oblique parameters are the products of renormalization under the SM, and thus warrant examination to ascertain whether such a quantity remains well-defined in the B-LSSM. A significant indicator is whether the divergence of oblique parameters is cancelled. From a technical standpoint, it appears that the requisite divergent elimination is guaranteed by the symmetry of the group, since $Tr[T^3Y]=0$.
\begin{table}[ht]
\centering
\caption{The charges and the mixed products of each superfield in the B-LSSM}
\label{TYB}
\renewcommand{\arraystretch}{1.5}
\setlength{\tabcolsep}{3pt}
\small
\begin{tabular}{|c|c|c|c|c|c|c|c|c|c|c|}
    \hline
    & $Q$ & $L$ & $H_d$ & $H_u$ & $d$ & $u$ & $e$ & $\nu$ & $\eta$ & $\bar{\eta}$ \\
    \hline
    $Y$ & $\frac{1}{6}$ & $-\frac{1}{2}$ & $-\frac{1}{2}$ & $\frac{1}{2}$ & $\frac{1}{3}$ & $\frac{2}{3}$ & $-1$ & $0$ & $0$ & $0$ \\
    \hline
    $T^3$ &
    $\left(\begin{smallmatrix} \frac{1}{2} & 0 \\ 0 & -\frac{1}{2} \end{smallmatrix}\right)$ &
    $\left(\begin{smallmatrix} \frac{1}{2} & 0 \\ 0 & -\frac{1}{2} \end{smallmatrix}\right)$ &
    $\left(\begin{smallmatrix} \frac{1}{2} & 0 \\ 0 & -\frac{1}{2} \end{smallmatrix}\right)$ &
    $\left(\begin{smallmatrix} \frac{1}{2} & 0 \\ 0 & -\frac{1}{2} \end{smallmatrix}\right)$ &
    $0$ & $0$ & $0$ & $0$ & $0$ & $0$ \\
    \hline
    $Y_B$ & $\frac{1}{6}$ & $-\frac{1}{2}$ & $0$ & $0$ & $-\frac{1}{6}$ & $-\frac{1}{6}$ & $\frac{1}{2}$ & $\frac{1}{2}$ & $-1$ & $1$ \\
    \hline
    $T^3 \times Y$ &
    $\left(\begin{smallmatrix} \frac{1}{12} & 0 \\ 0 & -\frac{1}{12} \end{smallmatrix}\right)$ &
    $\left(\begin{smallmatrix} -\frac{1}{2} & 0 \\ 0 & \frac{1}{2} \end{smallmatrix}\right)$ &
    $\left(\begin{smallmatrix} -\frac{1}{2} & 0 \\ 0 & \frac{1}{2} \end{smallmatrix}\right)$ &
    $\left(\begin{smallmatrix} \frac{1}{2} & 0 \\ 0 & -\frac{1}{2} \end{smallmatrix}\right)$ &
    $0$ & $0$ & $0$ & $0$ & $0$ & $0$ \\
    \hline
    $T^3 \times Y_B$ &
    $\left(\begin{smallmatrix} \frac{1}{12} & 0 \\ 0 & -\frac{1}{12} \end{smallmatrix}\right)$ &
    $\left(\begin{smallmatrix} -\frac{1}{2} & 0 \\ 0 & \frac{1}{2} \end{smallmatrix}\right)$ &
    $0$ & $0$ & $0$ & $0$ & $0$ & $0$ & $0$ & $0$ \\
    \hline
    $Y \times Y_B$ & $\frac{1}{36}$ & $\frac{1}{4}$ & $0$ & $0$ & $\frac{1}{18}$ & $-\frac{1}{9}$ & $-\frac{1}{2}$ & $0$ & $0$ & $0$ \\
    \hline
\end{tabular}
\end{table}
As can be observed in the Tab. \ref{TYB}, the self-energy functions, namely, the unrenormalized vacuum polarisation functions with coupling constants factored out, namely, the ($\Pi's$), will cancel out any divergence after traversing all the superfields. It is therefore reasonable to continue using the original definition of Peskin and Takeuchi\cite{Peskin:1991sw}:

\begin{align}
\alpha S\equiv&4e^2[\Pi'_{33}(0)-\Pi'_{3Q}(0)],\nonumber\\
\alpha T\equiv&\frac{e^2}{s_W^2c_W^2\widetilde{m_Z}^2}[\Pi_{11}(0)-\Pi_{33}(0)]\label{stu},\\
\alpha U\equiv&4e^2[\Pi'_{11}(0)-\Pi'_{31}(0)],\nonumber
\end{align}
where $\widetilde{m_Z}^2={c'}^2m_Z^2+{s'}^2m_{Z'}^2$.

In accordance with Peskin's approach, the one-particle-irreducible (1PI) self-energies of the photon, $Z$ and $Z'$ in the B-LSSM are expressed as a linear combination of the $T_3-Q-B$-mixed eigenstate self-energy.

\begin{subequations}
\begin{eqnarray}
&&\Pi_{AA}=e^2\Pi_{QQ}\label{pi1}.\\
&&\Pi_{ZA}=\frac{e^2c'}{s_Wc_W}(\Pi_{3Q}-s_W^2\Pi_{QQ})+es'g_E\Pi_{QB}\label{pi2}.\\
&&\Pi_{Z^{'}A}=-\frac{e^2s'}{s_Wc_W}(\Pi_{3Q}-s_W^2\Pi_{QQ})+ec'g_E\Pi_{QB}\label{pi3}.\\
&&\Pi_{ZZ}=\frac{e^2{c'}^2}{s_W^2c_W^2}(\Pi_{33}-2s_W^2\Pi_{3Q}+s_W^4\Pi_{QQ})\nonumber\\
&&\qquad\quad\;+2\frac{es'c'}{s_Wc_W}g_E(\Pi_{3B}-s^2_W\Pi_{QB})+{s'}^2g^2_E\Pi_{BB}\label{pi4}.\\
&&\Pi_{Z^{'}Z^{'}}=\frac{e^2{s'}^2}{s_W^2c_W^2}(\Pi_{33}-2s_W^2\Pi_{3Q}+s_W^4\Pi_{QQ})\nonumber\\
&&\qquad\quad\;-2\frac{es'c'}{s_Wc_W}g_E(\Pi_{3B}-s^2_W\Pi_{QB})+{c'}^2g^2_E\Pi_{BB}\label{pi5}.\\
&&\Pi_{ZZ^{'}}=-\frac{e^2c's'}{s_W^2c_W^2}(\Pi_{33}-2s_W^2\Pi_{3Q}+s_W^4\Pi_{QQ})\nonumber\\
&&\qquad\quad\;+2\frac{e({c'}^2-{s'}^2)}{s_Wc_W}g_E(\Pi_{3B}-s^2_W\Pi_{QB})+c's'g^2_E\Pi_{BB}\label{pi6}.
\end{eqnarray}
\end{subequations}
Where $g_E=g_B+\frac{Y}{Y_B}g_{YB}$, $Q$ indicates the electric charge, $T_3$ indicates the charge of $W_3$ and $B$ indicates the charge of $U(1)_{B-L}$ group. The 1PI self energy of $W^\pm$ is given as:
\begin{eqnarray}
\Pi_{WW}=\frac{e^2}{s_W^2}\Pi_{11}.
\label{ww}
\end{eqnarray}

We can linearly combine the above equations and define effective self-energy $\widetilde{\Pi}_{ZA}$ and $\widetilde{\Pi}_{ZZ}$ as follows to eliminate the self-energys "containing" $U_1$ charge $B$: $\widetilde{\Pi}_{ZA}=c'(\ref{pi2})-s'(\ref{pi3})$, $\widetilde{\Pi}_{ZZ}={c'}^2(\ref{pi4})-2c's'(\ref{pi6})+{s'}^2(\ref{pi5})$. Combining with (\ref{pi1}), one can obtain:
\begin{subequations}
\begin{eqnarray}
&&\Pi_{AA}=e^2\Pi_{QQ} \label{pi7}.\\
&&\widetilde{\Pi}_{ZA}=\frac{e^2}{s_Wc_W}(\Pi_{3Q}-s_W^2\Pi_{QQ}) \label{pi8}.\\
&&\widetilde{\Pi}_{ZZ}=\frac{e^2}{s_W^2c_W^2}(\Pi_{33}-2s_W^2\Pi_{3Q}+s_W^4\Pi_{QQ}) \label{pi9}.
\end{eqnarray}
\label{pi}
\end{subequations}
Combining the definition of oblique parameters Eq. (\ref{stu}), Eq. (\ref{ww}) and Eq. (\ref{pi}), one can obtain the expression of oblique parameters under the B-LSSM:
\begin{align}
\alpha S=&4s_Wc_W[\widetilde{\Pi}'_{ZZ}(0)-\frac{c_W^2-s_W^2}{s_Wc_W}\widetilde{\Pi}'_{Z\gamma}(0)-\Pi'_{\gamma\gamma}(0)],\nonumber\\
\alpha T=&\frac{\Pi_{WW}(0)}{m^2_W}-\frac{\widetilde{\Pi}_{ZZ}(0)+\frac{2s_W}{c_W}\widetilde{\Pi}_{Z\gamma}(0)+\frac{s^2_W}{c_W^2}\Pi'_{\gamma\gamma}(0)}{m^2_Z}\label{stum},\\
\alpha U=&4s_W^2[\Pi'_{WW}(0)-c_W^2\widetilde{\Pi}'_{ZZ}(0)-2s_wc_W\widetilde{\Pi}'_{Z\gamma}(0)-s_W^2\Pi'_{\gamma\gamma}(0)].\nonumber
\end{align}

\subsection{The S T U paramaters in the effective-Lagrangian\label{tree}}
For the case of $Z$ couples with two fermions, neutral-current of $Z$ particle can be  expressed as follows:
\begin{align}
J_Z^\mu=&[\frac{e}{s_Wc_W}(T^3-s^2Q)c^{'}+g_E Y_B s^{'}]\bar{\psi}\gamma^\mu P_L\psi\nonumber\\
&+[\frac{e}{s_Wc_W}(-s_W^2Q)c^{'}+g_E Y_B s^{'}]\bar{\psi}\gamma^\mu P_R\psi.\nonumber\\
\end{align}
Using Eq. (\ref{gb}) and $g_E=g_B+\frac{Y}{Y_B}g_{YB}$, the neutral-current coupling between the Z boson and SM leptons in the B-LSSM can be written as follows:
\begin{align}
\mathcal{L}_{N,Zff}=&G(c'-s'\frac{c's'(m^2_Z-m^2_{Z'})}{{c'}^2m^2_Z+{s'}^2m^2_{Z'}})\nonumber\\
&\times Z^{\mu}\bar{f}\gamma_{\mu}[T^3P_L-(s_W^2-\frac{s'(c^2_Wc's'(m^2_Z-m^2_{Z'})+\frac{xm_Zm_{Z'}}{2})}{{c'}^2m^2_Z+{s'}^2m^2_{Z'}})]f.
\label{nc}
\end{align}
Current experimental constraints on the $Z'$ boson impose a lower bound on the ratio $M_{Z'}/g_B > 6~\text{TeV}$ \cite{Cacciapaglia:2006pk,Carena:2004xs}. Theoretical analyses indicate that direct contributions from $Z'$-mediated neutral currents to electroweak-scale observables are suppressed to a precision level below $\mathcal{O}(10^{-3})$, as determined by the interplay of the $Z'$-boson mass scale and its coupling strength. Consequently, these neutral current effects can be safely neglected in calculations of electroweak precision parameters within the current experimental resolution. This suppression arises dominantly from the decoupling behavior of the heavy $Z'$, whose contributions to low-energy processes scale as $\sim (M_Z^2/M_{Z'}^2) \times g_B^2$. The stringent experimental bound on $M_{Z'}/g_B$ ensures that such terms remain subleading compared to SM electroweak radiative corrections. Using the effective-Lagrangian techniques given by \cite{Burgess:1993vc}
\begin{align}
\mathcal{L}_{N,Zee}=&\frac{e}{\hat{s_W}\hat{c_W}}(1+\frac{\alpha T}{2})Z^{\mu}\bar{f}\gamma_{\mu}[T^3P_L-(\hat{s_W}^2+\frac{\alpha S}{4(\hat{c_W}^2-\hat{s_W}^2)}-\frac{\hat{c_W}^2\hat{s_W}^2\alpha T}{\hat{c_W}^2-\hat{s_W}^2})]f,\nonumber\\
\mathcal{L}_{C,We\nu}=&-\frac{e}{\sqrt{2}\hat{s_W}}(1-\frac{\alpha S}{4(\hat{c_W}^2-\hat{s_W}^2)}+\frac{\hat{c_W}^2\alpha T}{2(\hat{c_W}^2-\hat{s_W}^2)}+\frac{\alpha U}{8\hat{s}^2}){W^{\pm}}^{\mu}\bar{f}\gamma_{\mu}P_Lf.
\label{nnc}
\end{align}
Where $\hat{s_W}$ and $\hat{c_W}$ are defined by
\begin{eqnarray}
\hat{s_W}\hat{c_W}m_Z=\frac{1}{2}e v=s_W c_W m_Z^{SM}.
\end{eqnarray}
Comparing Eq. (\ref{nc}) and Eq. (\ref{nnc}), one can obtain:
\begin{align}
\alpha T=&2[\frac{\hat{s_W}\hat{c_W}}{s_Wc_W}(c'-s'\frac{c's'(m^2_{Z'}-m^2_Z)}{{c'}^2m_Z^2+{s'}^2m_{Z'}^2})-1],\nonumber\\
\alpha S=&4\hat{c_W}^2\hat{s_W}^2\alpha T+4(\hat{c_W}^2-\hat{s_W}^2)(s_W^2-\hat{s_W}^2-\frac{s'}{c'}\frac{c_W^2c's'(m^2_{Z'}-m^2_Z)+\frac{xm_Zm_{Z'}}{2}}{{c'}^2m^2_Z+{s'}^2m_{Z'}^2}),\nonumber\\
\alpha U=&8\hat{s_W}^2(\frac{\hat{s_W}}{s_W}-1+\frac{\alpha S}{4(\hat{c_W}^2-\hat{s_W}^2)}-\frac{\hat{c_W}^2\alpha T}{2(\hat{c_W}^2-\hat{s_W}^2)}).
\label{stuo}
\end{align}
It can be demonstrated that the definition of $\hat{c_W}$ is indeed equivalent to the the Sirlin definition \cite{Sirlin:1980nh}, based on the values of $m_W$ and $m_Z$. In comparison to the intrinsic definition of $c_W$ at the tree level, these can be connected by means of the equation:
\begin{eqnarray}
&&\Delta s=\hat{s_W}-s_W,\nonumber\\
&&\hat{c_W}^2=\frac{m_W^2}{m_Z^2},\nonumber\\
&&c^2_W=\frac{m^2_W}{{c'}^2m_Z+{s'}^2m_{Z'}^2}=1-s_W^2\nonumber\\
&&\qquad\!\!=1-\hat{s_W}^2+2\hat{s_W}\Delta s=\hat{c_W}^2+2\hat{s_W}\Delta s.
\label{cw}
\end{eqnarray}
By comparing the three equations in Eq. (\ref{cw}), one can obtain:
\begin{align}
{s'}^2(\frac{m^2_{Z'}}{m^2_Z}-1)=-2\frac{\hat{s_W}}{\hat{c_W}^2}\Delta s\label{dssp}.
\end{align}
Combining Eq. (40) with Eq. (42), the following results can be derived through a perturbative analysis:
\begin{align}
\alpha T=&2\frac{\Delta s}{\hat{s_W}}-{s'}^2.\label{to}\\
\alpha S+\alpha U=&8\hat{s_W}\Delta s-4s'\frac{\hat{c_W}\hat{s_W}}{e}g_B.\label{l1}\\
2\hat{s_W}^2\alpha S+(\hat{s_W}^2-\hat{c_W}^2)\alpha U=&4\hat{c_W}^2\hat{s_W}^2\alpha T+8\hat{s_W}(\hat{s_W}^2-\hat{c_W}^2)\Delta s.\label{l2}
\end{align}
The aforementioned formulas are conducive to the analysis of relationship between $s'$, $\Delta s$ and $g_B$.

\subsection{Gauge-invariant electroweak interaction parameters\label{gstu}}
We have already provided the pinch contributions of all self energy-graphs of gauge particles in Eq. (\ref{pw}) and Eq. (\ref{NCP}), and also provided specific expressions for the $S$ $T$ $U$ parameters in (\ref{pi}) and Eq. (\ref{stum}). Combining the above four equations, we can obtain pinch parts of the $S$ $T$ $U$ paramaters. The gauge-invariant $S$ $T$ $U$ parameters can be represented as the sum of the pinch parts and the conventional parts:
\begin{align}
\alpha (S)_{GI}=&\alpha (S)_C+8g^2_2s_Wc_WI'_{WW}(0)({c'}^2m_Z^2+{s'}^2m_{Z'}^2).\\
\alpha (T)_{GI}=&\alpha (T)_C+4g^2_2[s_W^2I_{W\gamma}(0)+c_W^2({c'}^2I_{WZ}(0)+{s'}^2I_{WZ'}(0))-I_{WW}(0)].\\
\alpha (U)_{GI}=&\alpha (U)_C+16s_W^2g^2_2\{m^2_W[s_W^2I'_{W\gamma}(0)+c_W^2({c'}^2I'_{WZ}(0)+{s'}^2I'_{WZ'}(0))\nonumber\\
&-I'_{WW}(0)]-[s_W^2I_{W\gamma}(0)+c_W^2({c'}^2I_{WZ}(0)+{s'}^2I_{WZ'}(0))-I_{WW}(0)]\}.
\end{align}
The subscript C here follows the definition in the literature \cite{Degrassi:1993kn}, the $S$, $T$ and $U$ parameters exclude the pinch terms that ensure gauge independence.Formally speaking, our results are extensions of Degrassi's results \cite{Degrassi:1993kn} based on the SM. The divergences in both expressions cancel out. This confirms the assertion made in the first subsection of this section.
\section{The divergence cancellation of S T U parameters in the Neutralino-Chargino sector and squark sector\label{dcns}}
\indent\indent

In the context of one-loop diagram calculations, particularly in the context of renormalization, the elimination of divergences serves as a crucial indicator for the verification of calculation results. This section presents the divergence cancellation of the S, T, and U parameters in the neutralino-chargino and squark sectors, with the objective of further validating the theoretical framework.

\subsection{Neutralino-Chargino sector\label{ncs}}

\begin{figure}
\begin{minipage}[c]{1\textwidth}
\centering
\includegraphics[width=6.0in]{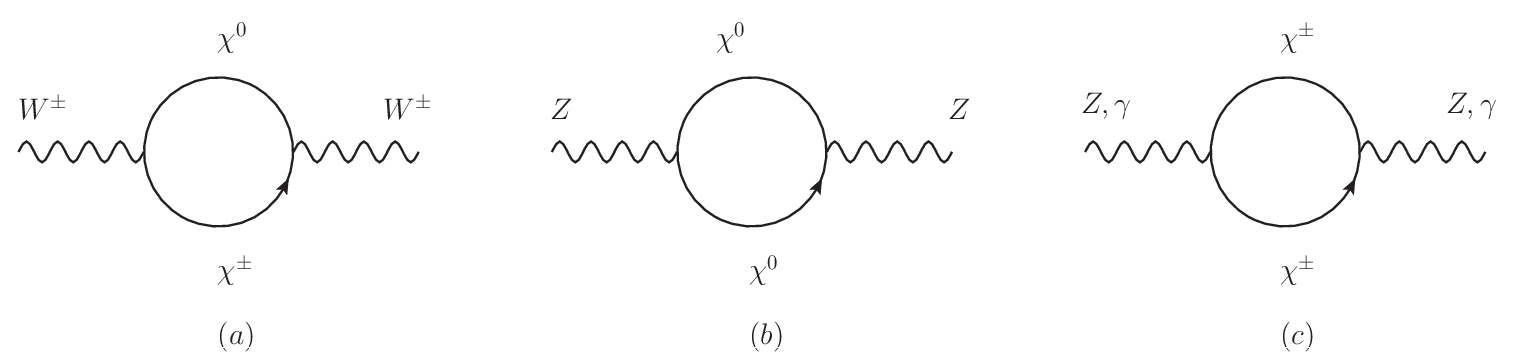}
\end{minipage}%
\caption[]{Vacuum polarization diagram with Neutralino-Chargino sector}
\label{fed}
\end{figure}
The calculation of the vacuum polarization diagrams of fermions as loop particles is presented in Appendix \ref{sfc}, along with a processing of all the coefficients involved. As a consequence of the non-divergence of the derivative of the $B_0$ function, the divergence of the $S$ and $U$ parameters is directly proportional to $\mathcal{C}_1$. By combining equations Eq. (\ref{pi}) and Appendix \ref{sfc}, the divergences of the $S$ $T$ and $U$ parameters ($div_S$, $div_T$ and $div_U$) can be expressed as:
\begin{align}
div_S\sim&-\frac{G^2}{4}[-\sum_{i,j}^2(U^*_{j2}U_{i2}U^*_{i2}U_{j2}+V^*_{j2}V_{i2}V^*_{i2}V_{j2})+\sum_{i,j}^7(N^*_{j3}N_{i3}-N^*_{j4}N_{i4})^2]=0.\nonumber\\
div_U\sim&-\frac{c_W^2G^2}{4}\{\sum_{i,j}^{7,2}[4U^*_{j1}N_{i2}N^*_{i2}U_{j1}+2U^*_{j2}N_{i3}N^*_{i3}U_{j2}+2\sqrt{2}(U^*_{j1}N_{i2}N^*_{i3}U_{j1}+h.c.)\nonumber\\
&+4V^*_{j1}N_{i2}N^*_{i2}V_{j1}+2V^*_{j2}N_{i4}N^*_{i4}V_{j2}-2\sqrt{2}(V^*_{j1}N_{i2}N^*_{i4}V_{j2}+h.c.)]\nonumber\\
&-\sum_{i,j}^2[4U^*_{j1}U_{i1}U_{i1}U^*_{j1}+4V^*_{j1}V_{i1}V_{i1}V^*_{j1}+U^*_{j2}U_{i2}U_{i2}U^*_{j2}+V^*_{j1}V_{i1}V_{i2}V^*_{j2}\nonumber\\
&+2c_W^2(U^*_{j1}U_{i1}U^*_{i2}U_{j2}+V^*_{j1}V_{i1}V^*_{i2}V_{j2})+2(1+2s_W^2)(U^*_{j2}U_{i2}U^*_{i1}U_{j1}+V^*_{j2}V_{i2}V^*_{i1}V_{j1})]\nonumber\\
&-\sum_{i,j}^7(N^*_{j3}N_{i3}-N^*_{j4}N_{i4})^2\}=0.\nonumber\\
div_T\sim&-\frac{G^2}{4m_Z^2}\{\mathcal{A}_1-\mathcal{A}_2\}.\nonumber
\end{align}
Where,
\begin{align}
\mathcal{A}_1=&\sum _{i,j}^{2,7}(m_{\chi_i^{\pm}}^2+m_{\chi_j^0}^2)[4U^*_{j1}N_{i2}N^*_{i2}U_{j1}+2U^*_{j2}N_{i3}N^*_{i3}U_{j2}+2\sqrt{2}(U^*_{j1}N_{i2}N^*_{i3}U_{j1}+h.c.)\nonumber\\
&+4V^*_{j1}N_{i2}N^*_{i2}V_{j1}+2V^*_{j2}N_{i4}N^*_{i4}V_{j2}-2\sqrt{2}(V^*_{j1}N_{i2}N^*_{i4}V_{j2}+h.c.)]\nonumber\\
&-\sum _{i,j}^2(m_{\chi_i^{\pm}}^2+m_{\chi_j^{\pm}}^2)[4c_W^4U^*_{j1}U_{i1}U^*_{i1}U_{j1}+c_{2W}^2U^*_{j2}U_{i2}U^*_{i2}U_{j2}+2c_W^2c_{2W}(U^*_{j1}U_{i1}U^*_{i2}U_{j2}+h.c.)\nonumber\\
&+4c_W^4V^*_{j1}V_{i1}V^*_{i1}V_{j1}+c_{2W}^22V^*_{j2}V_{i2}V^*_{i2}V_{j2}+2c_W^2c_{2W}(V^*_{j1}V_{i1}V^*_{i2}V_{j2}+h.c.)]\nonumber\\
&-\sum _{i,j}^7(m_{\chi_i^0}^2+m_{\chi_j^0}^2)(N^*_{j3}N_{i3}-N^*_{j4}N_{i4})^2.\nonumber
\end{align}
\begin{align}
\mathcal{A}_2=&\sum _{i,j}^{2,7}2m_{\chi_i^{\pm}}m_{\chi_j^0}[4U^*_{j1}N_{i2}V^*_{j1}N_{i2}-2U^*_{j2}N_{i3}V^*_{j2}N_{i4}-2\sqrt{2}U^*_{j1}N_{i2}V^*_{j2}N^*_{j4}+2\sqrt{2}U^*_{j2}N_{i3}V_{i1}N^*_{j2}\nonumber\\
&+4N^*_{j2}U_{1}N^*_{j2}V_{i1}+2N^*_{j3}U_{i2}N^*_{j4}V_{i2}-2\sqrt{2}N^*_{j2}U_{i1}N^*_{j4}V_{i2}+2\sqrt{2}N^*_{j3}U_{i2}N^*_{j2}V_{i1}]\nonumber\\
&-\sum_{i,j}^22m_{\chi_i^{\pm}}m_{\chi_j^{\pm}}[4c_W^4U^*_{j1}U_{i1}V_{j1}V^*_{i1}+c_{2W}^2U^*_{j2}U_{i2}V_{j2}V^*_{i2}+2c_W^2c_{2W}(U^*_{j1}U_{i1}V_{j2}V^*_{i2}+h.c.)\nonumber\\
&+4c_W^4V^*_{j1}V_{i1}U_{j1}U^*_{i2}+c_{2W}^2V^*_{j2}V_{i2}U_{j2}U^*_{i2}+2c_W^2c_{2W}(V^*_{j1}V_{i1}U_{j2}U^*_{i2}+h.c.)]\nonumber\\
&-\sum_{i,j}^72m_{\chi_i^0}m_{\chi_j^0}(N^*_{j3}N_{i3}-N^*_{j4}N_{i4})^2.\nonumber
\end{align}
Using the corresponding mass matrix diagonalization formula:
\begin{align}
&\sum_i^2U^*_{ik}m_{\chi^{\pm}_i}V_{ij}V^*_{ij}m_{\chi^{\pm}_i}U_{ik}=\sum_j^2(M_{\chi^{\pm}})^2_{kj},\nonumber\\
&\sum_{ij}^2U^*_{ik}U_{jk}V^*_{jl}V^*_{il}m_{\chi_i^{\pm}}m_{\chi_j^{\pm}}=(M_{\chi^{\pm}})^2_{kl}.
\end{align}
one can obtain:
\begin{align}
\mathcal{A}_1=&16M_2^2+g_2^2v^2+2\mu^2+2g_2^2v^2+\frac{1}{2}g_{YB}v^2-4c_W^4(4M_2^2+g_2^2v^2)\nonumber\\
&+c_{2W}^2(4\mu^2+g_2^2v^2)-(2\mu^2+2g_2^2v^2+\frac{1}{2}g_{YB}v^2)\nonumber\\
=&(16c_W^4-16)M_2^2+(4c_{2W}^2-4)\mu^2+(4c_W^4+c_{2W}^2-5)g_2^2v^2,\nonumber\\
\mathcal{A}_2=&2\times[8M^2_2+4\mu^2+g_2^2v^2-(8c_W^4M_2^2+2c_{2W}^2\mu^2+2c_W^2c_{2W}{g_2^2v^2})+2\mu^2]\nonumber\\
=&(16c_W^4-16)M_2^2+(4c_{2W}^2-4)\mu^2+(4c_W^4+c_{2W}^2-5)g_2^2v^2.\nonumber
\end{align}
It can be seen from this that their divergence cancels out each other !

\subsection{Squark sector\label{ss}}
When squarks are loop particles in the one-loop self-energy diagrams, the cancellation of divergences in the $S$, $T$, and $U$ parameters can be demonstrated as follows.
\begin{align}
div_S\sim&-\frac{4}{3\alpha}s_Wc_W[\mathcal{C}_a^{\gamma\gamma}-\frac{c_W^2-s_W^2}{s_Wc_W}\widetilde{\mathcal{C}}_a^{Z\gamma}-\widetilde{\mathcal{C}}_a^{ZZ}]\nonumber\\
=&\frac{4}{\alpha}s_Wc_W[(\sum^3_{k=1}Z^{U*}_{ik}Z^U_{ik})^2+4(\sum^6_{k=4}Z^{U*}_{ik}Z^U_{ik})^2-(\sum^3_{k=1}Z^{D*}_{ik}Z^D_{ik})^2+2(\sum^6_{k=4}Z^{D*}_{ik}Z^D_{ik})^2]\nonumber\\
=&0.\nonumber\\
div_U\sim&\frac{4}{\alpha}s_W^2[\mathcal{C}_a^{WW}-s_W^2\mathcal{C}_a^{\gamma\gamma}-2s_Wc_W\widetilde{\mathcal{C}}_a^{Z\gamma}-c_W^2\widetilde{\mathcal{C}}_a^{ZZ}]\nonumber\\
=&\frac{4}{\alpha}s_W^2[-\frac{1}{2}g^2_2\sum^3_{k=1}(Z^{D*}_{jk}Z^U_{ik})\sum^3_{l=1}(Z^{U*}_{il}Z^D_{jl})+\frac{1}{4}g^2_2(\sum^3_{k=1}Z^{U*}_{ik}Z^U_{ik})^2+\frac{1}{4}g^2_2(\sum^3_{k=1}Z^{D*}_{ik}Z^D_{ik})^2]\nonumber\\
=&0.\nonumber\\
div_T\sim&\frac{1}{m^2_W}[\sum^{6}_{i,j}\mathcal{C}_a^{WW}(m^2_{\widetilde{U}_i}+m^2_{\widetilde{D}_j})+\sum^{6}_{i}\mathcal{C}_b^{WWU}m^2_{\widetilde{U}_i}+\sum^{6}_{j}\mathcal{C}_b^{WWD}m^2_{\widetilde{D}_j}]\nonumber\\
&-\frac{1}{{c'}^2m^2_Z+{s'}^2m^2_{Z'}}[\sum^{6}_{i}\widetilde{\mathcal{C}}_a^{ZZU}m^2_{\widetilde{U}_i}+\sum^{6}_{j}\widetilde{\mathcal{C}}_a^{ZZD}m^2_{\widetilde{D}_j}+\frac{1}{2}\sum^{6}_{i}\widetilde{\mathcal{C}}_b^{ZZU}m^2_{\widetilde{U}_i}+\frac{1}{2}\sum^{6}_{j}\widetilde{\mathcal{C}}_b^{ZZD}m^2_{\widetilde{D}_j}\nonumber\\
&+\frac{s_W}{2c_W}(\sum^{6}_{i}\widetilde{\mathcal{C}}_a^{Z\gamma U}m^2_{\widetilde{U}_i}+\sum^{6}_{j}\widetilde{\mathcal{C}}_a^{Z\gamma D}m^2_{\widetilde{D}_j}+\frac{1}{2}\sum^{6}_{i}\widetilde{\mathcal{C}}_b^{Z\gamma U}m^2_{\widetilde{U}_i}+\frac{1}{2}\sum^{6}_{j}\widetilde{\mathcal{C}}_b^{Z\gamma D}m^2_{\widetilde{D}_j})]\nonumber\\
=&\frac{0}{m^2_W}-\frac{0}{{c'}^2m^2_Z+{s'}^2m^2_{Z'}}\nonumber\\
=&0.\nonumber
\end{align}
The specific coefficients for this process have been placed in Appendix \ref{IF}. In the demonstration above, we utilized the unitarity relation of the transposed matrix:
\begin{align}
\sum^6_{k=1}(Z^U_{ik}Z^{U*}_{jk})=\sum^6_{k=1}(Z^D_{ik}Z^{D*}_{jk})=\delta_{ij}.\nonumber
\end{align}

This work focuses on the neutralino-chargino and squark sectors of the B-LSSM, which are expected to provide the dominant new-physics contributions to electroweak precision observables, and within this framework we have constructed and verified the gauge-invariant formalism. Although the Higgs sector involves non-Abelian symmetry breaking and its mixing with Goldstone bosons, which introduces specific complexities in maintaining gauge invariance, its loop contributions to the $S$, $T$, and $U$ parameters have been shown to be extremely small already in the Standard Model. In the B-LSSM, the Higgs fields associated with the additional $U(1)$ breaking typically acquire masses well above the electroweak scale, further suppressing their contributions. Moreover, for the Higgs sector to yield a significant effect, it would generally require mixing between different generations of scalar fields, and such mixing is usually suppressed by the large mass scales. Therefore, at this stage of systematically studying the main sources of new-physics contributions, we have not included the Higgs sector in the detailed calculations.

\section{numerical analysis\label{NA}}

A global fit of the $S$, $T$, and $U$ parameters incorporating recent top quark mass measurements from CMS has been provided in Ref.\cite{ParticleDataGroup:2024cfk}:
\begin{align}
S=-0.04\pm0.10,\qquad &T=0.01\pm0.12,\qquad U=-0.01\pm0.09,\nonumber\\
\rho_{ST}=0.93,\qquad\quad\qquad&\rho_{SU}=-0.87,\quad\qquad\rho_{TU}=-0.70.\nonumber
\end{align}

In order to study how the parameters affect the vaules of the $S$, $T$ and $U$ parameters, we use the Mathematica to calculate the oblique parameters of the B-LSSM. Some parmeters we assumed are: $A_t=3~\rm TeV$, $A_b=3~\rm TeV$, $g_{YB}=-0.3$, $g_B=0.5$, $\tan\beta=20$, $\tan\beta'=1.5$, $m_{Z'}=4500~\rm GeV$, $\mu=200~\rm GeV$, $\mu_B=800~\rm GeV$, $m_B=600~\rm GeV$, $m_{BL}=800~\rm GeV$ $m_1=200~\rm GeV$ and $m_2=300~\rm GeV$. We use this set of data as a benchmark and run some of the variables respectively. Through a parameter scan, we can identify which model parameters are most sensitive to the $S$, $T$, and $U$ observables.

\begin{figure}
\begin{minipage}[c]{1\textwidth}
\centering
\includegraphics[width=3.0in]{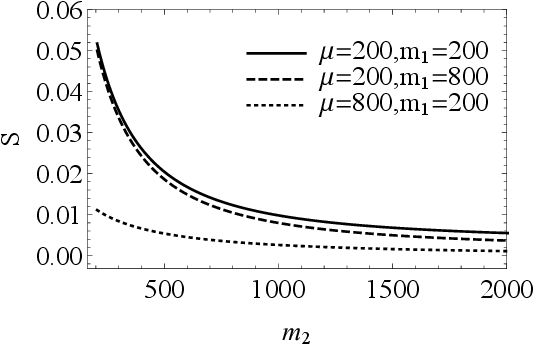}\includegraphics[width=3.2in]{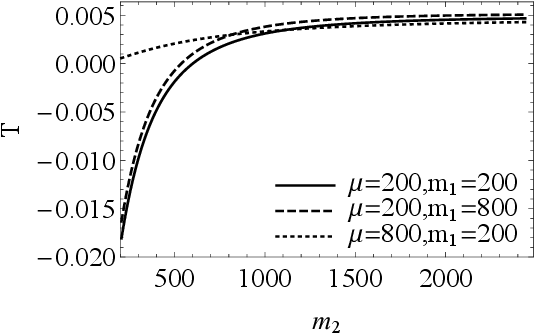}
\end{minipage}%
\caption[]{S and T parameters that are more sensitive to $m_1$, $m_2$, and $\mu$.}
\label{SM2}
\end{figure}
When we run $m_1$, $m_2$ and $\mu$ in the $(200~\rm GeV\sim2000~\rm GeV)$, $(200~\rm GeV\sim800~\rm GeV)$ and $(200~\rm GeV\sim800~\rm GeV)$ intervals respectively, we find that they are sensitive to the $S$ and $T$ parameters.
The Fig.~\ref{SM2} shows that the $S$ parameter gradually decreases with the increase of $m_1$, $m_2$, and $\mu$, while the $T$ parameter shows an upward trend with the increase of $m_1$ and $m_2$. $m_1$ and $m_2$ are the diagonal elements in the mass matrices of neutralinos and charginos, and their increase weakens the mixing of the matrices, which is the source of the contributions of oblique parameters. $m_2$ is the diagonal element with the highest value in the mass matrices of  neutralinos and charginos, which explains its sensitivity to oblique parameters relative to $m_1$.

\begin{figure}
\begin{minipage}[c]{1\textwidth}
\centering
\includegraphics[width=3.0in]{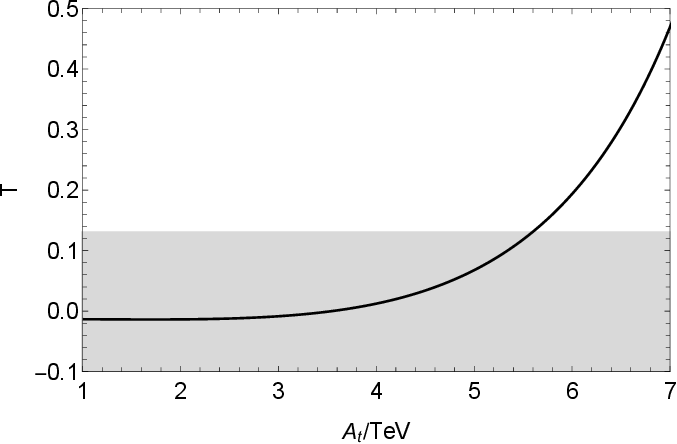}\includegraphics[width=3.0in]{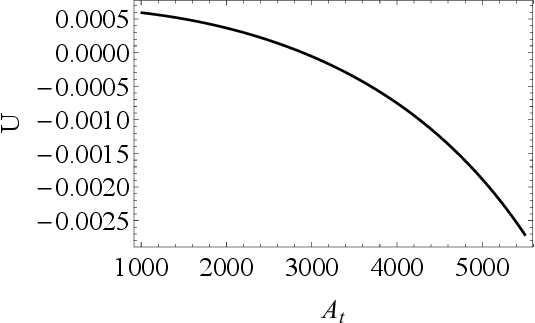}
\end{minipage}%
\caption[]{T and U parameters that are more sensitive to $A_t$.}
\label{At}
\end{figure}

Fig. \ref{At} illustrates the changes in the $T$ and $U$ parameters as the $A_t$ parameter varies from $1000~\rm GeV$ to $7000~\rm GeV$. The gray area indicates one standard deviation of the experimental observations. As shown, the $T$ parameter increases with the rise in $A_t$, while the $U$ parameter exhibits the opposite trend. The observed changes in the $T$ and $U$ parameters in Fig. \ref{At} are primarily due to the contributions from the Feynman diagrams involving squark fields as loop particles. The $A_t$ parameter, originating from the off-diagonal elements of the squark mass matrix, influences the mixing between the left-handed and right-handed $\widetilde{u}$ squarks ($\widetilde{u}_L$ and $\widetilde{u}_R$), which in turn drives the squark sector's contributions to the oblique parameters.

\begin{figure}
\begin{minipage}[c]{1\textwidth}
\centering
\includegraphics[width=6.5in]{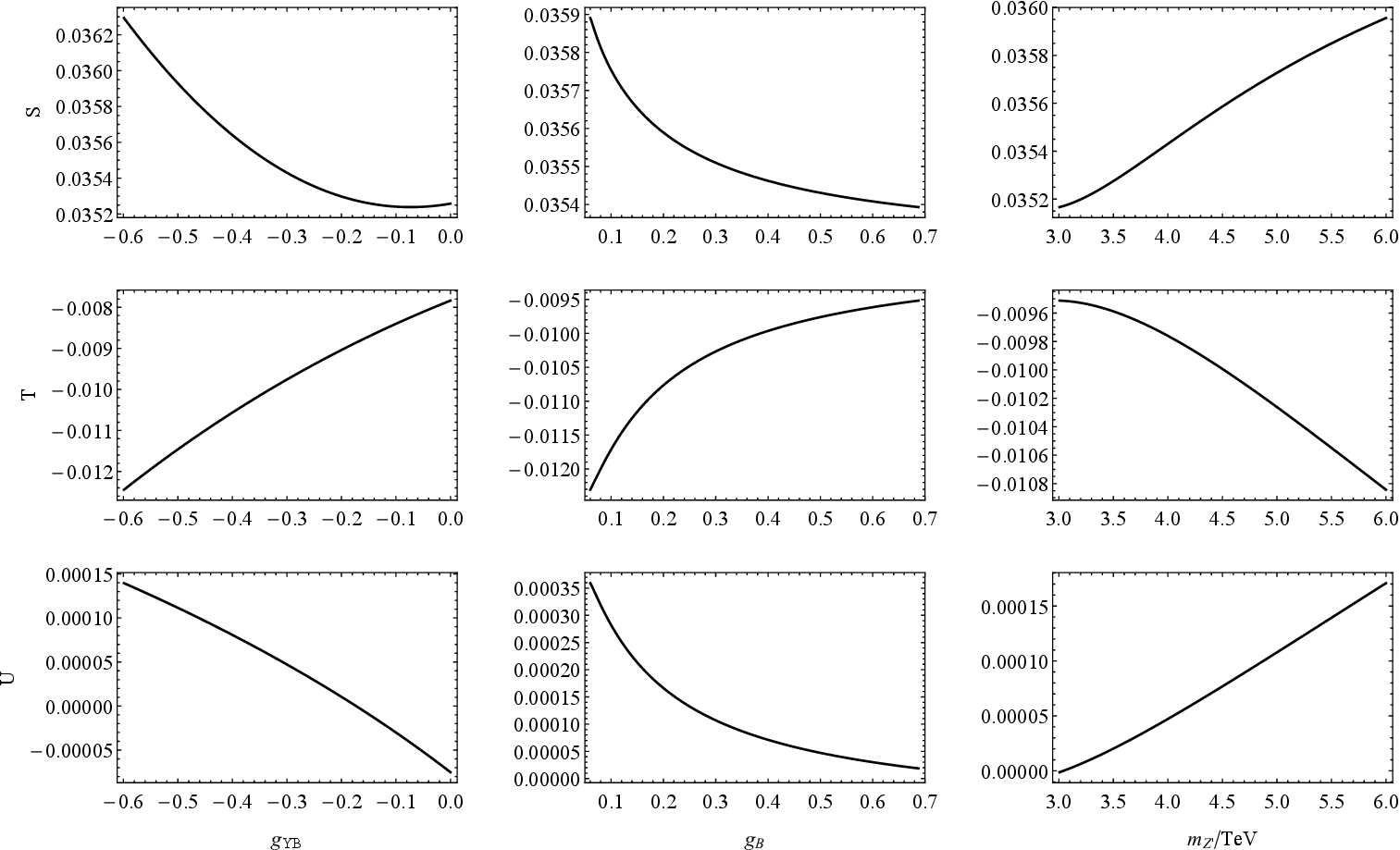}
\end{minipage}%
\caption[]{The dependence of the oblique parameters $S, T$, and $U$ on the B-LSSM distinctive parameters $g_{YB}$, $g_B$, and $m_{Z'}$. The SUSY-breaking parameters are fixed at the benchmark point.}
\label{sum}
\end{figure}

To genuinely highlight the distinctive features of the B-LSSM as opposed to the MSSM, it is essential to illustrate how the extended gauge sector parameters influence the electroweak precision observables. In Fig.~\ref{sum}, we present the dependence of the $S$, $T$, and $U$ parameters on the unique B-LSSM parameters: the gauge kinetic mixing coupling $g_{YB}$, the $U(1)_{B-L}$ gauge coupling $g_B$, and the extra gauge boson mass $m_{Z'}$, while keeping other supersymmetric parameters fixed at the aforementioned benchmark point.

As shown in the panels of Fig.~\ref{sum}, varying $m_{Z'}$ from 3000 GeV to 6000 GeV leads to a smooth, monotonic modulation of the oblique parameters. Because the $Z'$ boson is heavy, its decoupling behavior ensures that the absolute variations remain strictly perturbative. Furthermore, the figures demonstrate the pronounced effects of the extra gauge couplings. While $g_B$ dictates the interaction strength of the $U(1)_{B-L}$ sector, the parameter $g_{YB}$ explicitly governs the gauge kinetic mixing between the $U(1)_Y$ and $U(1)_{B-L}$ groups. This kinetic mixing effect directly modifies the $Z-Z'$ mass matrix, systematically shifting the $T$ parameter closer to zero while slightly reducing the $S$ parameter. These results explicitly demonstrate that while the loop-level sparticle mass patterns set the baseline, the unique $U(1)_{B-L}$ extensions provide a finite and distinctive tree-level modulation.

\begin{figure}
\begin{minipage}[c]{1\textwidth}
\centering
\includegraphics[width=2.2in]{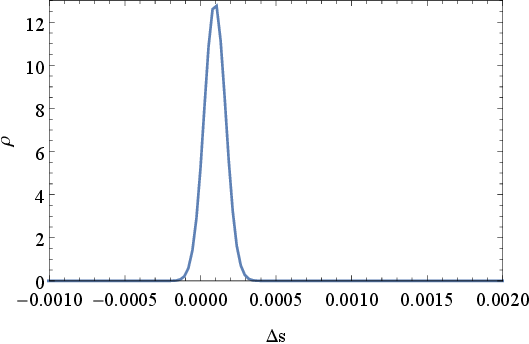}\includegraphics[width=2.2in]{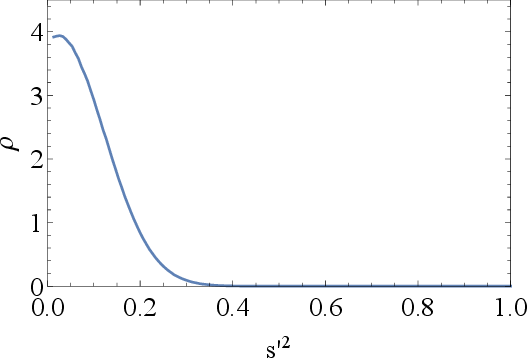}\includegraphics[width=2.2in]{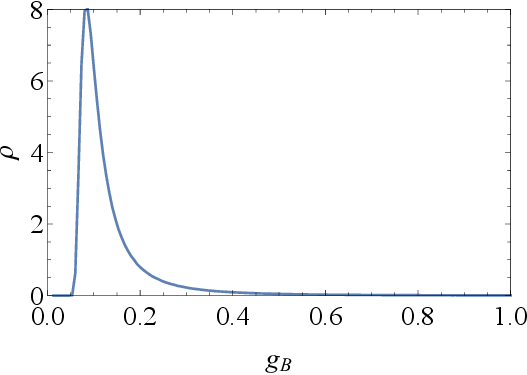}
\end{minipage}%
\caption[]{By using the Monte Carlo sampling method and generating statistical histograms, the probability distributions of the three parameters can be obtained.

}
\label{Sum}
\end{figure}

We performed a random three-dimensional sampling of the $S$, $T$, and $U$ parameters in Eq. (\ref{to}-\ref{l2}) based on the probability distribution results obtained from the experimental fits. Each randomly selected point corresponds to a unique set of values for \( \Delta s \), \( s'^2 \), and \( g_B \). We then conducted statistical analysis on the resulting \( \Delta s \), \( s'^2 \), and \( g_B \) values to generate probability histograms. As shown in Fig. \ref{Sum}, the distribution of \( \Delta s \) does not exceed one standard deviation of the current experimental measurements, making it challenging to extract further useful information from the measurement of the Weinberg angle. Additionally, the probability distribution of \( g_B \) exhibits a sharp peak around \( g_B = 0.08 \), with a rapid decay for \( g_B > 0.05 \). Analysis reveals that the probability for \( 0.05 < g_B < 0.2 \) exceeds ninety percent when compared to the entire range.

\section{Conclusions\label{CON}}
\indent\indent
Using the pinch technique, we derive the gauge-invariant self-energy of gauge bosons in the B-LSSM and obtain gauge-invariant oblique parameters. This approach ensures gauge symmetry when extracting information from S-matrix elements.

In models featuring an additional $U(1)$ symmetry compare to the Standard Model, the definitions of oblique parameters remain applicable. From the perspective of divergence cancellation, these definitions are comprehensive and can be extended to the models, which include various types of group expansion, thereby simplifying the complex definitions and proofs associated with the $S$, $T$, and $U$ parameters.

In this work, the contributions to the $S$, $T$, and $U$ parameters from two key sectors of the B-LSSM are calculated and the impacts of model parameters in the B-LSSM are analyzed. The parameter space of the B-LSSM we selected aligns well with the global fitting of the \( S \), \( T \), and \( U \) parameters reported in the experiments.

By comparing the effective Lagrangian with the model Lagrangian and incorporating constraints from experimental fits, the coupling constant of the additional $U(1)$ group can be related to the $S$, $T$, and $U$ parameters. This analysis allows us to derive effective constraints on these coefficients.

Finally, comparing the loop-level supersymmetric contributions with the tree-level extra gauge effects reveals a distinct physical hierarchy within the B-LSSM. Under current experimental precision, the globally fitted electroweak observables impose significantly more stringent constraints on the additional $U(1)_{B-L}$ gauge parameters (e.g., $g_B$ and $m_{Z'}$) than on the detailed supersymmetric mass spectrum. The latter, entering primarily via loop corrections, is naturally suppressed by the decoupling of heavy sparticles. In contrast, the extra gauge sector directly modulates the observables through tree-level kinetic mixing and mass matrix deformations. Consequently, contemporary precision measurements act as a powerful and direct probe of the extended gauge sector, validating the necessity of investigating the distinctive gauge extensions beyond the minimal supersymmetric framework.
\begin{acknowledgments}
\indent\indent
The work was supported by the Natural Science
Foundation of Guangxi Autonomous Region with Grant No. 2022GXNSFDA035068.The Natural Science Foundation of Hebei Province under Grant No.A2022104001,the Foundation of Baoding University under Grant No. 2023Z01
\end{acknowledgments}

\newpage
\appendix

\section{Integral formulas for some Feynman diagrams\label{IF}}

We provide the integral formulas for the vacuum polarization diagrams, expressing them uniformly in terms of the standard Passarino-Veltman A and B functions. We define $A$ fuctions:
\begin{align}
i\pi^2A_0(m)=&\mu^{4-n}\int d^nq\frac{1}{q^2+m^2-i\epsilon},\nonumber\\
i\pi^2A_{\mu}(m)=&\mu^{4-n}\int d^nq\frac{q_{\mu}}{q^2+m^2-i\epsilon}=0,\nonumber\\
i\pi^2A_{\mu\nu}(m)=&\mu^{4-n}\int d^nq\frac{q_{\mu}q_{\nu}}{q^2+m^2-i\epsilon}=i\pi^2A_2(m)g_{\mu\nu},
\end{align}
and $B$ fuctions:
\begin{align}
i\pi^2B_0(p^2;m_1,m_2)=&\mu^{4-n}\int d^nq\frac{1}{d_0d_1},\nonumber\\
i\pi^2B_{\mu}(p^2;m_1,m_2)=&\mu^{4-n}\int d^nq\frac{q_{\mu}}{d_0d_1}=i\pi^2B_1(p^2;m_1,m_2)p_{\mu},\nonumber\\
i\pi^2B_{\mu\nu}(p^2;m_1,m_2)=&\mu^{4-n}\int d^nq\frac{q_{\mu}q_{\nu}}{d_0d_1}=i\pi^2[B_{21}(p^2;m_1,m_2)p_{\mu}p_{\nu}+B_{22}(p^2;m_1,m_2)g_{\mu\nu}].
\end{align}
Where $d_0=q^2+m_1^2-i\epsilon$ and $d_1=(q+p)^2+m_2^2-i\epsilon$. We can list the divergence coefficients of these functions (The coefficients of $\frac{2}{\epsilon}-\gamma_E-\ln{\pi}$):
\begin{align}
&A_0(m)\sim-m^2,& &A_2(m)\sim\frac{m^4}{4},& &B_0(p^2;m_1,m_2)\sim1,&\nonumber\\
&B_1(p^2;m_1,m_2)\sim-\frac{1}{2},& &B_{21}(p^2;m_1,m_2)\sim\frac{1}{3},& &B_{22}(p^2;m_1,m_2)\sim-\frac{m^2_1+m_2^2}{4}-\frac{p^2}{12}.&\nonumber
\end{align}

\begin{figure}
\begin{minipage}[c]{1\textwidth}
\centering
\includegraphics[width=5.0in]{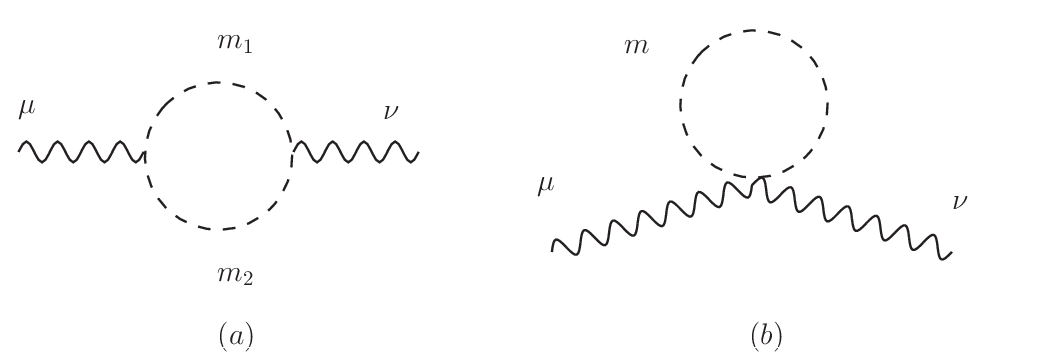}
\end{minipage}%
\caption[]{Feynman diagrams with scalar loop-particle}
\label{bed}
\end{figure}

Using the above formulas, the amplitudes of Fig. (\ref{bed}) can be represented as
\begin{align}
\mathcal{M}_a=&-\mathcal{C}_a\{4B_{22}(p^2;m_1,m_2)g_{\mu\nu}+[B_0+4B_1+4B_{21}](p^2;m_1,m_2)p_{\mu}p_{\nu}\},\nonumber\\
\mathcal{M}_b=&\mathcal{C}_bA_0(m)g_{\mu\nu}.
\end{align}
$\mathcal{C}_a$ and $\mathcal{C}_b$ are the coefficients from the vertices. We expand and sum the coefficients of the involved graphs.
\begin{align}
\mathcal{C}_a^{WW}=&-\frac{1}{2}g^2_2\sum^3_{k=1}(Z^{D*}_{ik}Z^U_{jk})\sum^3_{l=1}(Z^{U*}_{jl}Z^D_{il}),\nonumber\\
\mathcal{C}_b^{WWD}=&\frac{1}{2}g^2_2\sum^3_{k=1}(Z^{D*}_{ik}Z^D_{ik})\qquad\mathcal{C}_b^{WWU}=\frac{1}{2}g^2_2\sum^3_{k=1}(Z^{U*}_{jk}Z^U_{jk}),\nonumber\\
\mathcal{C}_a^{\gamma\gamma U}=&-\frac{4}{9}e^2(\sum_{k=1}^3Z^{U*}_{ik}Z^U_{jk}+\sum_{k=4}^6Z^{U*}_{ik}Z^U_{jk})^2,\nonumber\\
\mathcal{C}_a^{\gamma\gamma D}=&-\frac{1}{9}e^2(\sum_{k=1}^3Z^{D*}_{ik}Z^D_{jk}+\sum_{k=4}^6Z^{D*}_{ik}Z^D_{jk})^2,\nonumber\\
\mathcal{C}_b^{\gamma\gamma U}=&\frac{8}{9}e^2(\sum_{k=1}^3Z^{U*}_{ik}Z^U_{ik}+\sum_{k=4}^6Z^{U*}_{ik}Z^U_{ik}),\nonumber\\
\mathcal{C}_b^{\gamma\gamma D}=&\frac{1}{18}e^2(\sum_{k=1}^3Z^{D*}_{jk}Z^D_{jk}+\sum_{k=4}^6Z^{D*}_{jk}Z^D_{jk}),\nonumber\\
\widetilde{\mathcal{C}}_a^{ZZU}=&-\frac{1}{36}[G(1+2c_{2W})\sum^3_{k=1}Z^{U*}_{ik}Z^U_{jk}-2G(1-c_{2W})\sum^6_{k=4}Z^{U*}_{ik}Z^U_{jk}]^2,\nonumber\\
\widetilde{\mathcal{C}}_a^{ZZD}=&-\frac{1}{36}[G(-2-c_{2W})\sum^3_{k=1}Z^{D*}_{ik}Z^D_{jk}+G(1-c_{2W})\sum^6_{k=4}Z^{D*}_{ik}Z^D_{jk}]^2,\nonumber\\
\widetilde{\mathcal{C}}_b^{ZZU}=&\frac{1}{18}\{[G(1+2c_{2W})\sum^3_{k=1}Z^{U*}_{ik}Z^U_{ik}]^2+[2G(1-c_{2W})\sum^6_{k=4}Z^{U*}_{ik}Z^U_{ik}]^2\},\nonumber\\
\widetilde{\mathcal{C}}_b^{ZZD}=&\frac{1}{18}\{[G(-2-c_{2W})\sum^3_{k=1}Z^{D*}_{jk}Z^D_{jk}]^2+[G(1-c_{2W})\sum^6_{k=4}Z^{D*}_{jk}Z^D_{jk}]^2\},\nonumber\\
\widetilde{\mathcal{C}}_a^{Z\gamma U}=&-\frac{e}{9}(\sum_{k=1}^3Z^{U*}_{ik}Z^U_{jk}+\sum_{k=4}^6Z^{U*}_{ik}Z^U_{jk}) [G(1+2c_{2W})\sum^3_{k=1}Z^{U*}_{ik}Z^U_{jk}-2G(1-c_{2W})\sum^6_{k=4}Z^{U*}_{ik}Z^U_{jk}]^2,\nonumber\\
\widetilde{\mathcal{C}}_a^{Z\gamma D}=&-\frac{e}{18}(\sum_{k=1}^3Z^{D*}_{ik}Z^D_{jk}+\sum_{k=4}^6Z^{D*}_{ik}Z^D_{jk}) [G(-2-c_{2W})\sum^3_{k=1}Z^{D*}_{ik}Z^D_{jk}+G(1-c_{2W})\sum^6_{k=4}Z^{D*}_{ik}Z^D_{jk}]^2,\nonumber\\
\widetilde{\mathcal{C}}_b^{Z\gamma U}=&\frac{2}{9}e[G(1+2c_{2W})(\sum^3_{k=1}Z^{U*}_{ik}Z^U_{ik})^2-2G(1-c_{2W})(\sum^6_{k=4}Z^{U*}_{ik}Z^U_{ik})^2],\nonumber\\
\widetilde{\mathcal{C}}_b^{Z\gamma D}=&\frac{1}{9}e[G(-2-c_{2W})(\sum^3_{k=1}Z^{D*}_{jk}Z^D_{jk})^2+G(1-c_{2W})(\sum^6_{k=4}Z^{D*}_{jk}Z^D_{jk})^2].\nonumber
\end{align}
Where, we mark $c_{2W}=c^2_W-s^2_W=\cos{2\theta_W}$ and
\begin{align}
\widetilde{\mathcal{C}}^{ZZ}=&{c'}^2\mathcal{C}^{ZZ}-2c's'\mathcal{C}^{ZZ'}+{s'}^2\mathcal{C}^{Z'Z'},\nonumber\\
\widetilde{\mathcal{C}}^{Z\gamma}=&c'\mathcal{C}^{Z\gamma}-s'\mathcal{C}^{Z'\gamma},
\end{align}

\section{SELF-ENERGIES of Charginos and Neutralinos\label{sfc}}
For the convenience of batch processing, we divid the calculation results of graphs into two parts $\mathcal{M}_1$ and $\mathcal{M}_2$:
\begin{align}
\mathcal{M}=&-\int\frac{d^Dk}{(2\pi)^D}tr[A^{\mu}\frac{i}{/\!\!\!k-m_1}B^{\nu}\frac{i}{(/\!\!\!k+/\!\!\!q)-m_2}]\nonumber\\
&(A_LB_L+A_RB_R)\int\frac{d^Dk}{(2\pi)^D}\frac{1}{d_0d_1}[2k^{\mu}q^{\mu}+2k^{\nu}q^{\mu}+4k^{\mu}k^{\nu}-2(k^2+k\cdot q)g^{\mu\nu}]\nonumber\\
&+(A_LB_R+A_RB_L)\int\frac{d^Dk}{(2\pi)^D}\frac{2m_1m_2}{d_0d_1}g^{\mu\nu}.\nonumber
\end{align}
Where, $A^{\mu} = A_L P_L \gamma^{\mu} + A_R P_R \gamma^{\mu}$. We define
\begin{align}
\mathcal{M}_1=&\mathcal{C}_1\int\frac{d^Dk}{(2\pi)^D}\frac{1}{d_0d_1}[2k^{\mu}q^{\mu}+2k^{\nu}q^{\mu}+4k^{\mu}k^{\nu}-2(k^2+k\cdot q)g^{\mu\nu}]\nonumber\\
=&i\frac{1-\ln(2\pi)\epsilon}{16\pi^2}\mathcal{C}_1\{[4B_{22}(q^2)-2DB_{22}(q^2)-2B_{21}(q^2)q^2-2B_1(q^2)q^2]g^{\mu\nu}\nonumber\\&+4[B_1(q^2)+B_{21}(q^2)]q^{\mu}q^{\nu}\},\nonumber\\
\mathcal{M}_2=&\mathcal{C}_1\int\frac{d^Dk}{(2\pi)^D}\frac{2m_1m_2}{d_0d_1}g^{\mu\nu}\nonumber\\
=&i\frac{1-\ln(2\pi)\epsilon}{16\pi^2}\mathcal{C}_22m_1m_2B_0(q^2)g^{\mu\nu},\nonumber\\
\mathcal{C}_1=&(A_LB_L+A_RB_R),\nonumber\\ \mathcal{C}_2=&(A_LB_R+A_RB_L).
\end{align}
We can analyze their divergences:
\begin{align}
\mathcal{M}_1(q^2=0)\sim&\frac{i}{16\pi}\mathcal{C}_1(m_1^2+m_2^2)g^{\mu\nu},\nonumber\\
\mathcal{M}_2(q^2=0)\sim&2\frac{i}{16\pi}\mathcal{C}_2m_1m_2g^{\mu\nu},\nonumber\\
\frac{d\mathcal{M}(q^2=0)}{q^2}\sim&\frac{i}{16\pi}\frac{2}{3}\mathcal{C}_1g^{\mu\nu}.
\end{align}
We expand and sum the coefficients of the involved graphs.
\begin{align}
\mathcal{C}_1^{WW}=&-\frac{c_W^2 G^2}{4}[4U^*_{j1}N_{i2}N^*_{i2}U_{j1}+2U^*_{j2}N_{i3}N^*_{i3}U_{j2}+2\sqrt{2}(U^*_{j1}N_{i2}N^*_{i3}U_{j1}+h.c.)\nonumber\\
&+4V^*_{j1}N_{i2}N^*_{i2}V_{j1}+2V^*_{j2}N_{i4}N^*_{i4}V_{j2}-2\sqrt{2}(V^*_{j1}N_{i2}N^*_{i4}V_{j2}+h.c.)],\nonumber\\
\mathcal{C}_2^{WW}=&-\frac{c_W^2 G^2}{4}[4U^*_{j1}N_{i2}V^*_{j1}N_{i2}-2U^*_{j2}N_{i3}V^*_{j2}N_{i4}-2\sqrt{2}U^*_{j1}N_{i2}V^*_{j2}N^*_{j4}+2\sqrt{2}U^*_{j2}N_{i3}V_{i1}N^*_{j2}\nonumber\\
&+4N^*_{j2}U_{1}N^*_{j2}V_{i1}+2N^*_{j3}U_{i2}N^*_{j4}V_{i2}-2\sqrt{2}N^*_{j2}U_{i1}N^*_{j4}V_{i2}+2\sqrt{2}N^*_{j3}U_{i2}N^*_{j2}V_{i1}],\nonumber\\
\mathcal{C}_1^{\gamma\gamma}=&-\frac{e^2}{4}[4U^*_{j1}U_{i1}U^*_{i1}U_{j1}+4U^*_{j2}U_{i2}U^*_{i2}U_{j2}+4(U^*_{j1}U_{i1}U^*_{i2}U_{j2}+h.c.)\nonumber\\
&+4V^*_{j1}V_{i1}V^*_{i1}V_{j1}+4V^*_{j2}V_{i2}V^*_{i2}V_{j2}+4(V^*_{j1}V_{i1}V^*_{i2}V_{j2}+h.c.)],\nonumber\\
\mathcal{C}_2^{\gamma\gamma}=&-\frac{e^2}{4}[4U^*_{j1}U_{i1}V_{j1}V^*_{i1}+4U^*_{j2}U_{i2}V_{j2}V^*_{i2}+4U^*_{j1}U_{i1}V_{j2}V^*_{i2}+4U^*_{j2}U_{i2}V_{j1}V^*_{i1}\nonumber\\
&+4V^*_{j1}V_{i1}U_{j1}U^*_{i1}+4V^*_{j2}V_{i2}U_{j2}U^*_{i2}+4V^*_{j1}V_{i1}U_{j2}U^*_{i2}+4V^*_{j2}V_{i2}U_{j1}U^*_{i1}],\nonumber\\
\widetilde{\mathcal{C}}_{1\chi^{\pm}}^{ZZ}=&-\frac{G^2}{4}[4c_W^4U^*_{j1}U_{i1}U^*_{i1}U_{j1}+c_{2W}^2U^*_{j2}U_{i2}U^*_{i2}U_{j2}+2c_W^2c_{2W}(U^*_{j1}U_{i1}U^*_{i2}U_{j2}+h.c.)\nonumber\\
&+4c_W^4V^*_{j1}V_{i1}V^*_{i1}V_{j1}+c_{2W}^22V^*_{j2}V_{i2}V^*_{i2}V_{j2}+2c_W^2c_{2W}(V^*_{j1}V_{i1}V^*_{i2}V_{j2}+h.c.)],\nonumber\\
\widetilde{\mathcal{C}}_{2\chi^{\pm}}^{ZZ}=&-\frac{G^2}{4}[4c_W^4U^*_{j1}U_{i1}V_{j1}V^*_{i1}+c_{2W}^2U^*_{j2}U_{i2}V_{j2}V^*_{i2}+2c_W^2c_{2W}(U^*_{j1}U_{i1}V_{j2}V^*_{i2}+h.c.)\nonumber\\
&+4c_W^4V^*_{j1}V_{i1}U_{j1}U^*_{i2}+c_{2W}^2V^*_{j2}V_{i2}U_{j2}U^*_{i2}+2c_W^2c_{2W}(V^*_{j1}V_{i1}U_{j2}U^*_{i2}+h.c.)],\nonumber\\
\widetilde{\mathcal{C}}_{1\chi^0}^{ZZ}=&-2\frac{G^2}{4}(N^*_{j3}N_{i3}-N^*_{j4}N_{i4})^2,\nonumber\\
\widetilde{\mathcal{C}}_{2\chi^0}^{ZZ}=&2\frac{G^2}{4}(N^*_{j3}N_{i3}-N^*_{j4}N_{i4})^2,\nonumber\\
\widetilde{\mathcal{C}}_1^{Z\gamma}=&-\frac{eG}{4}[4c_W^2U^*_{j1}U_{i1}U^*_{i1}U_{j1}+2c_{2W}U^*_{j2}U_{i2}U^*_{i2}U_{j2}+2c_{2W}U^*_{j1}U_{i1}U^*_{i2}U_{j2}\nonumber\\
&+4c_W^2U^*_{j2}U_{i2}U^*_{i1}U_{j1}+4c_W^2V^*_{j1}V_{i1}V^*_{i1}V_{j1}+2c_{2W}V^*_{j2}V_{i2}V^*_{i2}V_{j2}\nonumber\\
&+2c_{2W}V^*_{j1}V_{i1}V^*_{i2}V_{j2}+4c_W^2V^*_{j2}V_{i2}V^*_{i1}V_{j1}],\nonumber\\
\widetilde{\mathcal{C}}_2^{Z\gamma}=&-\frac{eG}{4}[4c_W^2U^*_{j1}U_{i1}V_{j1}V^*_{i1}+2c_{2W}U^*_{j2}U_{i2}V_{i2}V^*_{j2}+2c_{2W}U^*_{j1}U_{i1}V_{j2}V^*_{i2}\nonumber\\
&+4c_W^2U^*_{j2}U_{i2}V_{j1}V^*_{i1}+4c_W^2V^*_{j1}V_{i1}U_{j1}U^*_{i1}+2c_{2W}V^*_{j2}V_{i2}U_{j2}U^*_{i2}\nonumber\\
&+2c_{2W}V^*_{j1}V_{i1}U_{j2}U^*_{i2}+4c_W^2V^*_{j2}V_{i2}U_{j1}U^*_{i1}].\nonumber
\label{cf}
\end{align}


\end{document}